\documentstyle[twocolumn,eqsecnum,aps,epic,eepic,epsfig]{revtex}

\begin{document}

\def\avg#1{\left< #1 \right>}
\def\abs#1{\left| #1 \right|}
\def\recip#1{\frac{1}{#1}}
\def\vhat#1{\hat{{\bf #1}}}
\def\smallfrac#1#2{{\textstyle\frac{#1}{#2}}}
\def\smallrecip#1{\smallfrac{1}{#1}}

\def\half{{1\over{2}}}
\def\Orabi{\Omega_{\rm rabi}}

\newcommand{\etal}{{\it et al.\ }}
\newcommand{\ibid}{{\it ibid.\ }}
\def\wx{\omega_x}
\def\wy{\omega_y}
\newcommand{\ofr}{({\bf r})}
\newcommand{\ofro}{({\bf r_0})}
\def\Eb{E_{\rm blue,rms}}
\def\Er{E_{\rm red,rms}}
\def\Es2{E_{0,{\rm sat}}^2}
\def\sb{s_{\rm blue}}
\def\sr{s_{\rm red}}

\draft

\title{Substrate-based atom waveguide using guided two-color evanescent
light fields}


\author{A.~H.~Barnett\cite{email}, S.~P.~Smith, M.~Olshanii,
K.~S.~Johnson\cite{kent},
A.~W.~Adams\cite{allan}, and M.~Prentiss}

\address{Lyman Laboratory, Harvard University, Cambridge,
Massachusetts 02138
}

\date{\today}
\maketitle


\begin{abstract}
We propose a dipole-force
linear waveguide which confines neutral
atoms up to $\lambda/2$ above a
microfabricated single-mode dielectric optical guide.
The optical guide
carries
far blue-detuned light in the horizontally-polarized TE mode
and
far red-detuned light in the vertically-polarized TM mode,
with both modes close to optical
cut-off.
A trapping minimum
in the transverse plane is formed above the optical guide
due to
the differing evanescent
decay lengths of the two modes.
This design allows
manufacture of
mechanically stable atom-optical elements on a
substrate.
We calculate the full vector bound modes for an arbitrary guide shape
using two-dimensional
non-uniform finite elements in the frequency-domain,
allowing us to
optimize atom waveguide properties.
We find that a rectangular optical guide of 0.8\,$\mu$m by 0.2\,$\mu$m
carrying 6\,mW of total laser power
(detuning
$\pm$15\,nm about the D2 line)
gives a trap depth of 
200\,$\mu$K for cesium atoms ($m_F = 0$),
transverse oscillation frequencies of $f_x = 40$\,kHz and
$f_y = 160$\,kHz,
collection area $\sim 1\,\mu$m$^{2}$
and coherence time
of 9\,ms.
We discuss the effects of
non-zero $m_F$, the D1 line, surface interactions, heating rate,
the substrate refractive index,
and the limits on waveguide bending radius.
\end{abstract}


\narrowtext

There has been much recent
progress
in the
trapping and cooling of neutral atoms, opening up new areas of
ultra-low energy and matter-wave physics \cite{nobelreviews}.
Waveguides for such atoms are of great interest for atom optics,
atom interferometery, and atom lithography.
Multimode atom waveguides act as
incoherent atom pipes
that could
trap atoms, transport them along complicated paths or
between different environments, or deliver
highly localized atom beams to a surface.
Single-mode waveguides (or multimode guides populated only by atoms in
the transverse
ground-state) could be used for coherent atom optics and interferometry
\cite{interf,adams}, as well as a tool for one-dimensional physics
such as boson-fermion duality
\cite{joseph,Maxim,1Deffects} and low-dimensional
Bose-Einstein condensation effects \cite{1Dbec}.

The optical
dipole-force has long been used to trap and manipulate
atoms \cite{nobelreviews} as well as
dielectric particles \cite{particles}.
The available intensity of lasers has allowed a multitude of such atom traps
in the far-detuned regime,
giving very low decoherence and heating rates,
and storage times on the order of
seconds \cite{grimm}.

Evanescent light waves have been popular in many
atom mirrors, traps and guides \cite{EWreview,oldmirrors}
since they can provide potentials with
high spatial gradients (decay lengths
$\sim \lambda / 2 \pi$ where $\lambda$ is the optical wavelength),
and use rigid
dielectric structures (prisms, fibers)
to define the potential shape.
For example,
there has been a series of repulsive (blue-detuned)
evanescent-wave (EW) traps which rely on gravity to
provide the counteracting force \cite{gravito}
and
recent
experiments have shown that hollow optical fibers can guide
atoms confined within the
hollow core using a repulsive evanescent field guided
by the fiber \cite{renn,ito}.

The idea of using an EW to provide both attractive and repulsive forces is due
to Ovchinnikov \etal \cite{ov91}, who proposed the use of two colors
({\it i.e.\ }red and blue detunings) and differing evanescent decay lengths
to achieve a trap with the potential
minimum a distance $\sim \lambda$ from a prism
surface.
Until now, this design has been restricted to planar traps (weak confinement
in the other two dimensions).

In this paper we discuss a two-color trap based on the
EW fields above a single-mode, submicron optical `channel' waveguide.
The trap provides
tight confinement in {\em two} dimensions
and allows free de Broglie wave propagation in the third, forming an
atomic waveguide that could transport atoms
a between $\lambda/4$ and $\lambda/2$ above the optical guide surface.
Our proposal is to utilize the
differing vertical
evanescent decay lengths of the two {\em polarizations} carried in the
single-mode optical guide (see Figure \ref{fig:geom}).
The physical origin of this decay length difference is
the fact that the TM mode is closer to optical {\em cut-off} than the TE mode at
the same frequency.

Our proposal is reminiscent of some existing resonant enhancement
schemes for EW mirrors
(demonstrated with surface plasmons \cite{esslinger} and
dielectric waveguides \cite{finesse}) but with a radical change from a
planar geometry to a linear, the mechanism for exciting the guide,
and the simultaneous guiding of a second frequency of opposite detuning.
It also shares the feature of two guided colors with an atom
trap proposal
using microsphere whispering-gallery modes\cite{whisper}.

Our design has many desirable
experimental features:
1) very little optical power is required to obtain large
trapping intensities since the optical bound mode has very small cross
sectional area ($\sim 0.3 \, \mu m^{2}$),
2) the optical field is {\em non-divergent}, so can
be maintained over distances
orders of magnitude further than diffraction-limited propagation
in free space
allows,
3) the trapping potential is well-known, mechanically stable,
and insensitive
to experimental parameters other than the optical powers,
since it is defined by single-mode intensity
distributions fixed relative to a substrate,
4) 
fabrication of
arrays of closely spaced atom waveguides is possible\cite{hank},
for
parallel lithography or measurement, creating ``on-chip'' integrated
atom-optical
elements,
5) the atoms are exposed providing additional optical and physical access
(a feature not shared with hollow-fiber designs), and
6)
the velocity of the atoms along
the direction of the waveguide could be controlled by standing waves in
the light carried by the
optical guide \cite{kawata}.

Compared to a
Zeeman-effect magnetic trap for neutral atoms,
far-detuned
optical dipole-force traps can have comparable trapping times,
but typically an order of magnitude less depth and transverse
mode spacings than recent magnetic traps \cite{magrev,micromagtraps,joseph}.
However, in microfabricated applications the stray
magnetic fields decay as a power law with distance,
whereas evanescent light fields decay exponentially (ignoring for now any
scattering into free space caused by optical defects). 
We believe this could give guided optical traps 
a distinct advantage in terms of achievable density of independent
atom-optical elements on a single substrate.

Also, optical traps have
the advantage that
there is no significant loss mechanism which can remove atoms
from the trap
(assuming the thermal energy is much less than the trap depth):
spontaneous events cause a small heating rate, and
non-adiabatic changes in $m_F$ can change the optical potential but
not the fact that the atom remains trapped.
This is to be contrasted with a non-adiabatic spin-flip
event in a magnetic trap, which
results in loss of the atom.
This makes optical waveguides particularly attractive for
incoherent transport, when the loss of coherence due to the
spontaneous events is unimportant.
Finally, optical manipulation has the advantage over magnetic manipulation
in terms of high possible switching speeds.

This paper is organized as follows.
In Section \ref{sec:concept}
we describe the dipole potential,
the exponential approximation for the EW fields, and the mechanism
for the difference in decay length.
We show how we optimized the optical guide dimensions,
in the case of a rectangular
guide on a substrate of unity refractive index (for $m_F = 0$),
and discuss some design objectives and implementation issues.
In Section \ref{sec:prop}
we give simulated results for cesium atoms:
trap depth, coherence time, transverse mode spacing and Q factor, and
spontaneous heating
rate.
We also show how depth and coherence time are generally limited by only
two parameters
(the detuning and the normalized decay length difference).
We study both
the case of a substrate refractive index of unity, and in Section
\ref{sec:sub}
the more realistic index of 1.32.
We describe the numerical electromagnetic finite element technique in
Section \ref{sec:num}, including the accuracy achieved.
Section \ref{sec:further} is an investigation of two
potential causes of loss or
decoherence of atoms, namely interactions with the dielectric surface and
bending of the waveguide.
Finally in Section \ref{sec:conc} we conclude and give some future
prospects for this proposal.

\section{Trap Concept}
\label{sec:concept}

\subsection{Theory of the light potential}

An atom in a near-resonant light field of frequency $\omega$
experiences both a
conservative force
(due to stimulated photon exchange) and a dissipative force (due
to spontaneous photon emission)\cite{gordon,ct-theory}.
The conservative force
is the gradient of a spatially-dependent potential $U_{\rm dip}\ofr$
which
can be viewed as the time-averaged induced
dipole interaction energy
(proportional to the real part of the classical polarizability)
in the electric
field,
or equally well as the `light shift' (that is, energy level shift due to
the ac Stark effect) of the atomic ground state \cite{adams,grimm}.

We assume that we apply a monochromatic light field of detuning
$\Delta \equiv \omega - \omega_0$ to an
alkali atom (with the $n\,s \rightarrow n\,p$ transition resonant at
$\omega_0$),
in the {\em far-detuned} regime
($\Delta$ greater
than the excited state hyperfine splitting, but much less than $\omega_0$)
and the {\em low saturation} regime
($\Orabi \ll \Delta$, where the Rabi flopping rate $\Orabi$
is defined \cite{definitions} by
$\hbar \Orabi \equiv \mu E_0$, the dipole matrix element multiplied by
the electric field amplitude).
The dipole potential has both a scalar and a magnetic part:
\begin{eqnarray}
\label{eq:alkali}
U_{\rm dip}\ofr & = &
\beta_s \, \frac{\hbar \Gamma}{8} \, \frac{\Gamma}{\Delta} \, s\ofr\
\nonumber \\
& + & \mu_{\rm Bohr}\, g_{\rm nuc}(L,S,J,i,F) m_F |{\bf H}\ofr|
\,,
\end{eqnarray}
where
$\Gamma$ is the spontaneous decay rate,
$s\ofr$ is the spatially-dependent saturation parameter,
and the potential is taken to be much less than the ground state
hyperfine splitting.
Only the magnetic part is affected by $m_F$, which is defined as
the projection of the total angular momentum $F$
on the direction of
the local effective magnetic field ${\bf H}\ofr$.
The constant $g_{\rm nuc}$
is the nuclear Land\'{e} g-factor appropriate for the $F$ of the ground state.
The scalar potential is identical to the case of a
two-level atom, apart from the strength
factor $\beta_s$ which is $\smallfrac{1}{3}$ for detuning from
the D1 line, $\smallfrac{2}{3}$
for the D2 \cite{grimm}.

It is important to be precise with the definition of the
saturation parameter.
We write
\begin{equation}
\label{eq:esat}
s\ofr \equiv \frac{E_0^2\ofr}{\Es2} \,,
\end{equation}
where $\Es2 \equiv 2 I_{\rm sat} / \epsilon_0 c$ (in the MKSA system)
is the squared
electric field amplitude in a plane wave of intensity $I_{\rm sat}$
\cite{definitions}.
This avoids the ambiguities present
with the usual definition $s\ofr \equiv I\ofr/I_{\rm sat}$
in the case of a general monochromatic light field
(composed of an arbitrary
coherent sum of travelling and evanescent waves),
and emphasizes that it is the local electric field alone that causes
the dipole potential.

The effective magnetic field has a strength and direction
given by the circularly polarized component of the electric field\cite{grimm},
which can be written thus:
\begin{equation}
\label{eq:mag}
\mu_{\rm Bohr} {\bf H}\ofr \, = \,
\beta_m \, \frac{\hbar \Gamma}{8} \, \frac{\Gamma}{\Delta} \,
\frac{\Re[{\bf E}_0^{*}\ofr] \times \Im[{\bf E}_0\ofr]}{\Es2}
\,,
\end{equation}
where the strength factor $\beta_m$ is $-\smallfrac{2}{3}$
for D1,
$\smallfrac{2}{3}$ for D2, and we take the physical electric field
(with amplitude $E_0\ofr \equiv |{\bf E}_0\ofr|$)
to be the real part of a complex field ${\bf E}({\bf r},t) \equiv
{\bf E}_0\ofr \exp(-i \omega t)$.
The reason for the `dummy' constant $\mu_{\rm Bohr}$ is aesthetic,
so that (\ref{eq:alkali}) can be expressed
in a standard magnetic form.
Note that for nonzero $m_F$ the magnetic and scalar
contributions to the potential are of similar order, if the
fields have significant circular polarization
(this will be true for our trapping fields, for the reason that the
optical guide is
close to cut-off).

The fact that $U_{\rm dip}\ofr$
has its sign controlled by
the sign of the detuning allows both attractive (red-detuned) and repulsive
(blue-detuned) potentials to be created.
The potential scales as $I/\Delta$ but the spontaneous emission
rate scales as $I/\Delta^2$; from this follows the well-known
result that,
if coherence time is an important factor,
it is best to
be far off-resonance and use high
intensities in order to achieve the
desired trap depth \cite{grimm,adams}.

For simplicity, in this paper we
will restrict our further analysis and simulations to $m_F = 0$, although
our initial
calculations suggest that the effect of the magnetic part of our potential when
trapping in other $m_F$ states will not pose major problems
(assuming the spin axis adiabatically follows the ${\bf H}\ofr$ field
direction),
and can even be
used to our advantage by increasing the depth and the
transverse oscillation frequency
$\omega_x$ in the case $m_F > 0$ \cite{future}.
%
Also, we will consider the effect of only a single resonance (choosing
D2 because it has a larger $\beta_s$ than D1),
which is a valid approximation when the detunings from this resonance
are much less than
the alkali atom fine structure splitting $\Delta_{\rm fs}$.
Even when it becomes advantageous to use a large detuning
of the order of $\Delta_{\rm fs}$,
it is possible to cancel the effect of the other resonance by a simple
shift in the two detunings
(as we will see at the end of Section~\ref{sec:dep}).

If we now have two light fields of differing frequency, the
atomic potentials add\cite{gordon,ov91}, as long as
we assume that the timescale of atomic motion is much slower than the
beating period (that is, the inverse
of the frequency difference).
In our case, atomic motion occurs at $\sim 10^{5}$\,Hz and our light field
frequency
difference is $\sim 10^{15}$\,Hz, so this assumption is valid.
Choosing equal but opposite detunings $\pm \Delta$
about the D2 line, the trapping
potential for $m_F = 0$ is
\begin{equation}
\label{eq:pot}
	U_{\rm dip}\ofr = \frac{2}{3} \,
	\frac{\hbar \Gamma}{8}  \, \frac{\Gamma}{\Delta} \,
	[\sb\ofr -\sr\ofr] \,,
\end{equation}
written in terms of the saturation parameters
for the two colors.
The spontaneous scattering rate\cite{adams}
is a factor $\Delta / \Gamma$ smaller than
$U_{\rm dip}\ofr / \hbar$ (in fact this relation applies
for any $m_F$ state), but it depends on the
{\em sum} of the saturation parameters rather than the
{\em difference}.
It is also spatially dependent, and has the form
\begin{equation}
\label{eq:spont}
	\Gamma_{\rm scatt}\ofr = \frac{2}{3} \,
	\frac{\Gamma}{8}  \left( \frac{\Gamma}{\Delta} \right)^{\!2} 
	[\sb\ofr + \sr\ofr] \,.
\end{equation}

\subsection{Design of the light fields}
\label{sec:des}

Our basic task is to create intense evanescent light fields with a potential
minimum sufficiently far
from a dielectric surface to make
the surface interaction potential and heating mechanisms
negligible (discussed in Section \ref{sec:surf}).
The main difficulty arises because the evanescent fields have
a typical exponential decay length $\sim \lambda / 2\pi$,
so if we are to have a trap of useful
depth, we are restricted to keep it within roughly $\lambda$ of the surface
(less than a micron).

\begin{figure}[t]
\centerline{\epsfig{figure=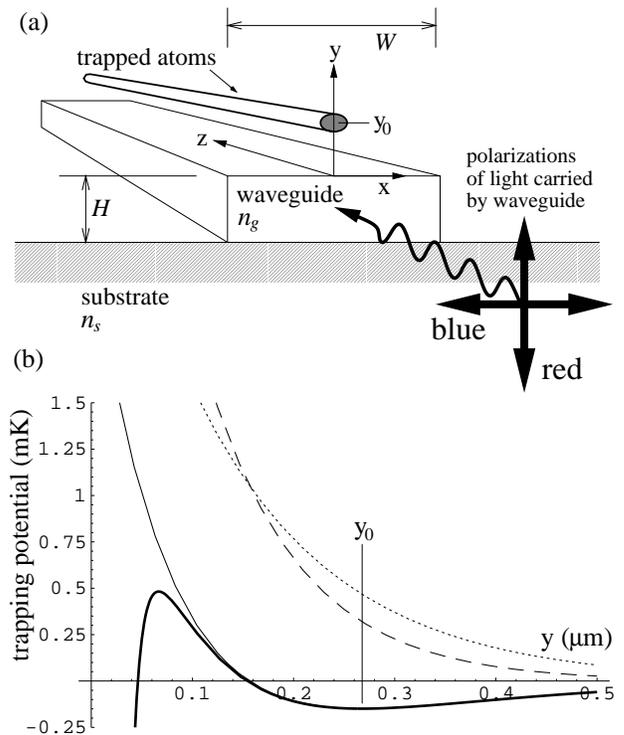,width=3.2in}}
\vspace{0.1in}
\caption{(a) Shows trap geometry, dielectric guide dimensions,
incoming laser polarizations, and the
cartesian axes;
(b) shows the trapping potential above the dielectric
along a vertical slice at $x=0$.
The component due to red-detuned light (absolute value shown as dotted line)
subtracts from that of blue-detuned light (dashed line)
to give the total dipole potential $U_{\rm dip}$
(thin solid line). This is modified by the Casimir surface interaction
(Section \ref{sec:surf}), giving the final potential
(thick line).
Here the trap depth of 150\,$\mu$K and coherence time of 12\,ms is
generated in
our design by 2 \,mW total guided laser power detuned by $\pm 15$\,nm from the
cesium D2 line.
}
\label{fig:geom}
\end{figure}

A potential minimum in one dimension can be obtained using a blue (repulsive)
light field of higher intensity at the
dielectric surface than the red (attractive) light field, and ensuring the
decay lengths obey $L_{\rm red} > L_{\rm blue}$, giving a potential
of the form
\begin{equation}
\label{eq:uofy}
	U_{\rm dip}(y) =
	A_{\rm blue} e^{-y/L_{\rm blue}} \,- A_{\rm red} e^{-y/L_{\rm red}}\,.
\end{equation}
This gives a repulsive force at short range, which becomes attractive at long
range (see Figure \ref{fig:geom}b),
and is the scheme for the planar trap of Ovchinnikov \etal\cite{ov91}.
A large amount of insight into our proposed trap can be gleaned from this
simple one-dimensional model (which we call the
{\em exponential approximation}),
because the squared electric fields above the guide will turn out to
approximate
exponential forms in the vertical direction quite closely.

If we define a normalized decay length difference
$\alpha_L \equiv (L_{\rm red} - L_{\rm blue})/L_{\rm blue}$, then
we can give two reasons why increasing $\alpha_L$ is a vital
design objective.
Firstly, it is easy to show that for small $\alpha_L$
the deepest available trap depth (found by optimizing the ratio of surface
intensities $A_{\rm red}/A_{\rm blue}$)
scales as $\alpha_L$.
Secondly, a larger $\alpha_L$ is beneficial for trap coherence,
(giving
a smaller spontaneous decay rate
at a given trap depth and detuning), because the sum of the
intensities can be kept lower (see equation (\ref{eq:spont})) for a given
intensity difference (equation (\ref{eq:pot})).
We will quantify this latter connection in Section \ref{sec:dep}.

Our two key differences from the proposal of Ovchinnikov \etal are as follows.
Firstly, we create a non-zero $\alpha_L$ by using two orthogonally-polarized
bound modes of a dielectric slab guide, which have different evanescent decay
lengths at the same frequency\cite{freqnote}.
This contrasts with Ovchinnikov \etal
who suggest varying the decay lengths
by varying the reflection angles from the inside surface of a glass prism.
Secondly, horizontal confinement is achieved by limiting the width of the
slab guide to approximately $\lambda$
(forming what is called a {\em channel guide} \cite{tamir}),
which automatically creates a
maximum in each light intensity field
in the horizontal direction. This results in a tight horizontal confinement
in the atomic potential, of similar size to the vertical confinement, and
is something very hard to achieve in a prism geometry.

A schematic of our design is shown in Figure \ref{fig:geom}a.
The optical guide
height $H$ and width $W$ are kept small enough to guarantee
that
there are exactly two bound modes, differing in
polarization but not in nodal structure
(in optical terminology this is called {\it single-mode}):
$E_{11}^x$ has an electric field predominantly 
in the x direction\cite{FLnote}, and is
to be excited by blue-detuned laser light,
and $E_{11}^y$ has electric field predominantly in the y direction
and is
to be excited by red-detuned laser light.
We can see
why their vertical decay lengths differ by considering the case of the slab
({\em i.e.} taking the width
$W \rightarrow \infty$), where these 
modes are simply the slab TE and TM modes respectively.
For both these slab modes the purely transverse field
obeys the differential equation
\begin{equation}
\label{eq:slab}
	\frac{\partial^2 \phi}{\partial y^2} =
	k_0^2 \, [(n_{\rm eff}^{(i)})^2 - n(y)^2] \, \phi \,,
\end{equation}
where $\phi = E_x$ ($H_x$) for the TE (TM) mode, the eigenvalue
$n_{\rm eff}^{(i)} \equiv k_z^{(i)}/k_0$ is the effective refractive index for
the $i^{\rm th}$ mode
($k_z$ being the wavenumber in the propagation direction
and
$k_0$ the free space wavenumber), and $n(y)$ is the spatially-dependent
refractive index \cite{tamir}.
(This is equivalent to a one-dimensional quantum problem in the
direction normal to the slab, in a potential $-k_0^2 n(y)^2$ with
$\hbar^2/2m = 1$).
However the boundary conditions on the slab surfaces differ for the two mode
types: $\phi$ is always continuous, but
$\partial \phi / \partial y$ is continuous for TE as opposed to
$n^{-2} \partial \phi / \partial y$ continuous for TM.
This asymmetry exists because the permittivity $\epsilon = n^2$ varies
in space but the permeability $\mu$ is assumed to be constant.
This discontinuity in the gradient for the TM mode forces
it to have a lower $n_{\rm eff}$ than that of the TE mode,
which means it is less tightly bound
so has a longer
evanescent decay length.
This effect becomes more pronounced as the slab index increases or as optical
cut-off is
approached (which
happens when $n_{\rm eff}$ decreases until it reaches $n_s$
and the mode becomes unbound).
This tendency
is preserved even as the width is decreased to only
a few times the height, as in our scheme.

No analytic solution exists for the general rectangular guide,
so we used the finite element method discussed in Section \ref{sec:num}
to solve for the bound mode $n_{\rm eff}$ values and fields as a
function of guide dimensions.
Figure \ref{fig:cutoff} shows the resulting `cut-off curves', that is, contours
of constant $n_{\rm eff}$ in the parameter space $(W,H)$.
In this example we chose a guide index $n_g = 1.56$
(typical for a polymer dielectric) and, as a preliminary case,
a substrate index $n_s = 1$.
%
%

The single-mode region, in which we wish to remain,
is bounded below by the $E^y_{11}$ and $E^x_{11}$ curves
and above by the $E^y_{21}$ curve.
Note that, as in any dielectric guiding structure uniform in the z axis, the
lowest two modes ($E^y_{11}$ and $E^x_{11}$ in our case)
{\em never} truly reach cut-off, rather, they approach it exponentially
as the guide cross-section is shrunk to zero.
For this reason, we chose the practical definition of cut-off for these modes
to be
$n_{\rm eff} = 1.05$, which corresponds to only about 20\% of the power being
carried inside the guide.
In contrast, higher modes do have true cut-offs\cite{snitzer,tamir}
(this distinction is illustrated
by the dispersion curves of Figure \ref{fig:num}), and for the $E^y_{21}$
mode our (numerically limited) contour choice of $n_{\rm eff} = 1.02$ falls
very close to the true cut-off curve.

Using the numerically calculated electric field strengths of the
$E^y_{11}$ and $E^x_{11}$
modes, we found the red and blue guided laser powers which gave the deepest
trap, subject to the constraint of fixed total power (keeping the
detuning constant \cite{detuningnote}).
We also imposed the restriction that the zero of trapping potential come no
closer than 100\,nm along the vertical line $x=0$, which kept the trap minimum
a reasonable distance from the surface (see Section
\ref{sec:surf}).
Performing this optimization over a region of the parameter space covering
the single-mode region gave a contour plot of maximum achievable depth for a
given total power, shown within the rectangle overlayed on Figure
\ref{fig:cutoff}.
This depth increases from negligible values in the top left to
the largest depths in the lower right, indicating that choosing $W$ and $H$ to
be in this latter corner of the single-mode region is best for depth.
The depth shows very little variation with $W$ in this corner, rather
it is clear that varying $H$ to stay within our definition of
the single-mode region
has become the limiting factor on achievable depth.
We indicate a practical choice of $W = 0.97 \,\lambda$ and $H = 0.25 \,\lambda$
as a small marker on Figure \ref{fig:cutoff}.
Example trapping potentials shapes possible with these parameters are
shown in Figure
\ref{fig:pot};
we discuss their properties in Section
\ref{sec:prop}.

In Figure \ref{fig:geom}a the direct
excitation of the optical guide by the two
laser beams is
shown only schematically.
In a realistic experimental setup this coupling
into the guide would happen on the order of
a centimeter
from the atom guiding region, and could involve tapered or Bragg
couplers\cite{tamir}
from beams or from other fibers.
At this distance we estimate that isotropic stray
light due to an insertion loss of 0.5
would have 8 orders of magnitude less intensity than
the EW fields in the guiding region.
Assuming the light is coherent, this limits the fractional
modulation of the guiding
potential to $10^{-4}$.
More improvements are possible, including the use of absorbing shields,
bending the guide through large angles
away from the original coupling direction, and reducing the coherence length.

\subsection{Discussion of optical cut-off and substrate choice}

We have shown
in Figure \ref{fig:cutoff} that parameters optimized for trap depth are near
optical cut-off.
It is worth gathering together the physical reasons for this.
Firstly, on general geometric grounds,
the typical available intensities in a guide scale inversely with the
effective cross-sectional area of the bound mode, which, far from cut-off
(that is, when $n_g - n_{\rm eff} \ll n_g - n_s$)
follows very closely the cross-sectional area of the guide.
This makes it favorable to shrink guiding structures to areas less than a
square wavelength, where they generally become single-mode.
Secondly, once we're in the single-mode regime, as we approach
cut-off the mode power is carried increasingly outside the guide, increasing
the ratio of surface intensity to guide center intensity.
Thirdly, the evanescent decay length in the vacuum is longer
as we approach cut-off
(recall that in the case of the slab, this is exactly expressed by
$L^{-1} = 2 k_0 \sqrt{n_{\rm eff}^2 - 1}$ \cite{tamir}, where the factor of 2
arises because we are considering intensity decay length rather than
amplitude).
In the special case of $n_s = 1$,
the decay length diverges to infinity as we approach
cut-off.
Finally,
the ratio $\alpha_L$ becomes larger as we approach cut-off, with corresponding
beneficial effects on depth and coherence (due to increasing the
`goodness factor' we will introduce in Section \ref{sec:dep}).

Unfortunately, these purely theoretical reasons for approaching cut-off
are in opposition to more practical ones.
The closer to cut-off a guide is, the more sensitive it is to manufacturing
variations in cross-section: in our case this will be predominantly a
sensitivity to $H$.
The result is that small variations in $H$
cause large variations in mode size, or, at worst, complete cut-off.
If the mode size change is rapid (nonadiabatic)
along the z axis, (for instance if
this change is due
to surface roughness or refractive index inhomogeneities)
the resulting mismatches will be
a source of scattering of
the guided power.
Any coherent scattering back down the guide will set up
periodic modulations of the light field over long distances. (One way to
reduce the distance over which coherent addition is possible
is to use very broad
line-width light sources, which would dramatically reduce
modulations due to both guided and stray scattered light).
The consequence for the atoms would be a z-dependent trap depth and shape,
and this could lead to partial reflection or even localization of
the matter waves.
In general we expect scattering
to
limit how close to cut-off we can reliably operate.

With regard to the substrate, further practical issues arise.
In the above cut-off calculation we chose the simplest
case of $n_s = 1$, corresponding to a guide surrounded by vacuum.
A real substrate with $n_s > 1$ has the
unfortunate effect of limiting the propagation constant $k_z$
of strictly bound modes
to be larger than the freely propagating wavevector in the substrate;
in other words, $n_{\rm eff} > n_s$ must hold or the light field
will rapidly tunnel into the
`attractive potential' of the
substrate.
This in turn limits the decay lengths and $\alpha_L$ that can be achieved.
The trap properties quoted in the abstract and in Sections
\ref{sec:dep} and \ref{sec:other} paper rely on very low
$n_{\rm eff}$ values (1.07 for the $E^y_{11}$ mode, 1.18 for $E^x_{11}$)
for the reason that a low $n_{\rm eff}$ is the only way to
create long evanescent decay lengths in the vacuum.
(This is equivalent to Ovchinnikov \etal
choosing reflection angles very close to
critical\cite{ov91}.)
However, in Section \ref{sec:sub} we present preliminary results for
a substrate of sodium fluoride (the lowest-index common optical
mineral, at $n = 1.32$), and do not believe the substrate alters
the basic feasibility of our waveguide.

For completeness, here we list some other possible
approaches to the substrate issue.
1) Use an aerogel substrate, which can have exceptionally low
refractive indices and low loss (films of several $\mu$m thickness with
indices of about 1.1 can be produced\cite{aerogels}).
2) Use a dielectric multilayer substrate
with an effective index
of unity or less (very low loss multilayer mirrors\cite{kimble}
with effective indices
less than unity
can be created).
3) Investigate if there exist guide shapes which have sufficiently small
tunnelling rate into a conventional substrate that the fact that the modes are
not strictly bound becomes irrelevant (for instance, a wedge shape with the
smallest face in contact with the substrate).
4) Unsupported guiding structures could
be produced over short distances\cite{hank}.
Finally, it is important to note that the idea of replacing the substrate by a
metallic reflective layer
is not practical because they are too lossy.

Ultimately, the best values of $W$ and $H$, the best guide
cross-sectional shape,
and the substrate
choice
will depend on many of the above
factors and is an area for further research.

\section{Trap properties}
\label{sec:prop}

In the bulk of
this Section we will examine the atomic waveguide
properties for light nearly resonant
with the D2 line of
cesium,
using an optical guide of index 1.56 of the
dimensions $W = 0.97 \,\lambda$ and $H = 0.25 \,\lambda$
from Section \ref{sec:des},
and a substrate of unity index.
The saturation intensity for cesium is 11.2\,W/m$^2$ \cite{adams}, and
its resonant wavelength of 852\,nm requires that the physical guide size is
0.83\,$\mu$m by 0.21\,$\mu$m.
(At the end of the Section we present preliminary calculations for a
$n_s =  1.32$ substrate and a different guide, and discuss
how the atom waveguide properties are changed).

\begin{figure}[t]
\centerline{\epsfig{figure=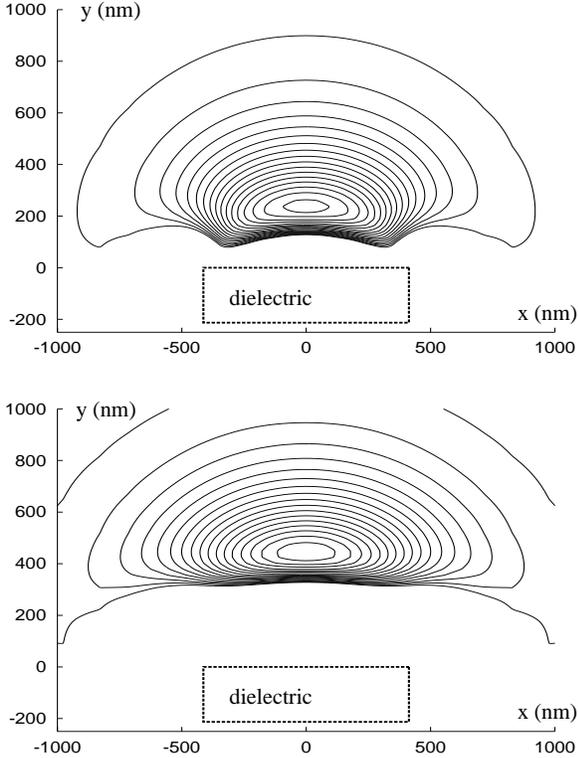,width=3.2in}}
\vspace{0.1in}
\caption{Trapping potential shapes in the xy plane, with guide
dimensions of 0.83\,$\mu$m by 0.21\,$\mu$m and an index of 1.56.
In each case the maximum depth is $100\,\mu$K and the
coherence time of atoms in
the ground state is 19\,ms, achieved with
detuning $\pm 15$\,nm from the cesium D2 line.
The outer contour shows a depth
of $3\,\mu$K, the cesium MOT temperature.
Subsequent contours are
spaced by $6\,\mu$K.
The plots illustrate the range of trapping distances acheivable:
the upper trap ($p=0.4$, using a total guided power of 1\,mW)
has a minimum 0.24\,$\mu$m from the surface; the lower trap ($p=0.2$,
total guided power of 7.5\,mW) has a minimum 0.44\,$\mu$m from the surface.}
\label{fig:pot}
\end{figure}

Given the guide, we are free to choose three experimental parameters, namely
the optical powers carried in the two modes,
and the detuning $\Delta$ (assumed to be
symmetric, that is, to be of equal magnitude for red and blue beams,
because little advantage can be gained with an unsymmetric detuning).
The first two of these can usefully be reexpressed as total power
$P_{\rm tot} \equiv P_{\rm red} + P_{\rm blue}$, and the power ratio
$p \equiv P_{\rm red} / P_{\rm blue}$.
The trap shape will be affected by $p$ alone: we show
the trapping potential shapes achievable at the two practical extremes of
$p=0.4$ and $p=0.2$
in Figure \ref{fig:pot}, where we have chosen $P_{\rm tot}$
and $\Delta$ to give
identical trap depths and coherence times.
Smaller $p$ values cause the trap minimum to move further from the surface
(a distinct advantage),
to be less ``bean'' shaped ({\it i.e.} to have smaller cubic deviations from
a 2D harmonic oscillator), and to cause a slight increase in collection area.
It is possible to achieve a trap minimum as distant as
$y_0 = 0.52\,\lambda$ from the
surface when $p=0.2$.
The only disadvantage to implementing these smaller $p$ values is that a higher
$P_{\rm tot}$ is required to achieve the same trap depth and
coherence time (for instance a factor of 7.5 increase is required as we take
$p$ from 0.4 to 0.2).
This can be quantified within the exponential approximation,
and it can be found that the
total power required to maintain a given depth and coherence time with a fixed
trap geometry scales as
$P_{\rm tot} \sim (1+p)/p^{1 + 1/\alpha_L}$.

If we were purely interested in maximizing trap depth at a given $P_{\rm tot}$
and detuning, it would be best to make $p$ as large as possible, however if we
take $p$ much larger than 0.4 the trap is brought so close that
the corners of the ``bean'' shape touch the dielectric surface
(see Figure \ref{fig:pot}, upper plot) and we will lose
effective collection area due to sticking of atoms onto this surface.

\subsection{Depth, coherence time, and Q factor}
\label{sec:dep}
We may ask what trade-offs are necessary between trap depth and coherence time.
It turns out that, within the exponential approximation
(\ref{eq:uofy}), this is elegantly quantifyable.
We can define a `goodness factor'
\begin{equation}
\label{eq:Gdef}
	G \equiv \frac{\sb\ofro - \sr\ofro}
	{\sb\ofro + \sr\ofro}
	= \frac{\Gamma}{\hbar |\Delta|} \, U_{\rm max} \tau_{\rm coh},
\end{equation}
where the trap minimum position ${\bf r_0}$
is at $(x=0, y=y_0)$, and the second equality is
verified by substitution of (\ref{eq:pot}) and (\ref{eq:spont}), and
defining $U_{\rm max} \equiv | U_{\rm dip}\ofro |$ and
$\tau_{\rm coh} \equiv \Gamma_{\rm scatt}^{-1}\ofro$.
We use this latter definition because we are interested in the coherence time
of
atoms spending time close to the trap minimum (which will certainly be
true for the transverse ground state.)
Using (\ref{eq:uofy})
to solve for $y_0$ and evaluate the `goodness factor',
it turns out that the factor is independent of either laser power ({\it i.e.} of
either $A_{\rm red}$ or $A_{\rm blue}$), giving
\begin{equation}
\label{eq:Geval}
	G = \frac{L_{\rm red}-L_{\rm blue}}{L_{\rm red}+L_{\rm blue}}
	= \frac{\alpha_L}{2 + \alpha_L} \, .
\end{equation}
Combining (\ref{eq:Gdef}) and (\ref{eq:Geval}) gives
\begin{equation}
\label{eq:tradeoff}
	U_{\rm max} \tau_{\rm coh} = \frac{\alpha_L}{2 + \alpha_L}
	\, \frac{\hbar |\Delta|}{\Gamma}\, ,
\end{equation}
fixing the product of achievable depth and coherence time as a constant
multiple of the detuning.
This is a remarkable result since it shows that increasing $\alpha_L$ is
really {\em the} only objective in the field design of two-color EW traps.
We can write this in units more convenient for cesium trap design, thus,
\begin{eqnarray}
\label{eq:expt}
	\frac{U_{\rm max}}{\mu {\rm K}} \cdot \frac{\tau_{\rm coh}}{{\rm ms}}
	& = &
	(644.2) \frac{\alpha_L}{2 + \alpha_L} \cdot \frac{|\Delta|}{{\rm nm}}
	\nonumber \\
	& = &
	(122) \cdot \frac{|\Delta|}{{\rm nm}} \, ,
\end{eqnarray}
where the value $\alpha_L = 0.47 \pm 0.02$ (taken from best-fit exponentials to the
numerically-found squared electric fields for the guide dimensions of Section
\ref{sec:des} with $n_s = 1$)
has been
substituted to give the the final form.
This design expression does not give the $P_{\rm tot}$ required to reach
a desired balance between $U_{\rm max}$ and $\tau_{\rm coh}$, however, the
total laser power is usually in the mW range, several orders of magnitude less
than in most free-space trap designs.
For instance, with $P_{\rm tot} = 20$\,mW, $p=0.4$ and $\Delta = \pm15$\,nm
we could generate a trap of 2\,mK depth
with the relatively short coherence time of 0.9\,ms.
The transverse oscillation frequencies in this trap would be
$\omega_x / 2\pi = 116$\,kHz and
$\omega_y / 2\pi = 490$\,kHz
(the field shapes fix this ratio at about 1:4),
giving an atomic mode spacing due to the x motion of 5.6\,$\mu$K, roughly twice
the cesium MOT temperature, and
a characteristic ground-state size of 26\,nm by 12\,nm.

For coherent guiding, we can define a more physically meaningful
figure of merit,
$Q \equiv \omega_{\bot} \tau_{\rm coh}$, which tells us the typical number of
coherent transverse oscillations we can expect multiplied by $2\pi$
({\it i.e.} it is the Q-factor of the transverse oscillations).
We should choose $\omega_{\bot} = \omega_x$ since this is
the smaller of the transverse frequencies in our case.
For $Q \gg 1$ the transverse atomic modes will be well resolved, and our guide
can be a useful interferometric device.
Using (\ref{eq:tradeoff}), in conjunction with the
fact that when the trapping potential shape is fixed then
$\omega_x$ is proportional to the square root of the depth,
tells us that for a given trap and detuning, $Q \propto 1/\omega_x$.
For a higher $Q$ we should choose smaller transverse oscillation frequencies,
that is, shallower traps.
For example, the 2\,mK trap discussed above has $Q \approx 650$, but
if we reduce it to a 20\,$\mu$K trap of the same $\Delta$ (by changing $p$ or
the laser powers), the $Q$ is 10 times larger.
Increasing $\Delta$ would allow even higher $Q$ to be realized.

The dependence on detuning in (\ref{eq:expt})
is another way of
expressing the advantages already known about using far off-resonant
beams\cite{adams,grimm}.
However, our single-resonance approximation will break down if the
detuning is too large: we have (somewhat arbitrarily) chosen
a detuning limit of 15\,nm, as compared to $\Delta_{\rm fs} = 43$\,nm
for cesium.
At this limit, the additional dipole potential created due to the 
detunings from the D1 line is very significant.
However, by removing the detuning symmetry
(changing $\Delta_{+}$ from +15\,nm to +12.07\,nm and
$\Delta_{-}$ from -15\,nm to -17.14\,nm), the desired
D2 single-resonance approximation
potential is recovered in the true physical situation of both resonances
present.
(These required shifts, which are of order $\Delta^2/\Delta_{\rm fs}$, can
easily be found using the expression for the sum of dipole potentials
from the two lines.)
An additional necessity for our limit is the fact that any larger detunings
start to demand separate bound-mode calculations for the two colors, a
treatment we reserve for the future.
This detuning limit in turn limits the depths,
coherence times and Q-factors we quote here,
but
we anticipate similar
future EW atom waveguide designs which explore the region
$|\Delta| > \Delta_{\rm fs}$ (or even $|\Delta| \sim \omega_0$),
and achieve much better coherence.
%

\subsection{Other properties}
\label{sec:other}

We estimate the collection area of the trap as the cross-sectional region
within which the potential is deeper than the typical cesium
MOT energy $k_B T_{\rm MOT}$.
For our example 100\,$\mu$K traps of Figure \ref{fig:pot} this area is about
1\,$\mu$m$^2$.
However, it is not possible to do much better than this with our design:
if
one tries to increase the area by increasing the overall trap depth,
the $k_B T_{\rm MOT}$
contour touches the substrate, indicating that atoms at this energy
can reach the substrate surface, where they will stick,
limiting the effective collection area.

To investigate the lifetime of atoms transported incoherently (the
multi-mode regime), we can calculate the heating rate
along similar lines as Grimm and Weidem\"{u}ller\cite{grimm}.
We start with their equation (23) which gives the rate of change of
the average of the total energy of atomic motion
$E = E_{\rm kin} + E_{\rm pot}$ as
\begin{equation}
\label{eq:heat}
	\dot{\overline{E}} = k_B T_R \, \overline{\Gamma_{\rm scatt}} \, ,
\end{equation}
$T_R$ being the recoil temperature,
and use the assumption that in an equilibrated 3D trap
$\overline{E_{\rm kin}} = \smallfrac{3}{2} k_B T$.
Since there is harmonic motion in two directions but free motion in the third,
the virial theorem gives us
$\overline{E_{\rm pot}} =  \smallfrac{2}{3} \overline{E_{\rm kin}}$.
Combining this with
$\overline{\Gamma_{\rm scatt}} = U_{\rm max} \Gamma / G \hbar \Delta$
from (\ref{eq:Gdef}) gives the heating rate
\begin{equation}
\label{eq:tdot}
	\dot{T} = \frac{2}{5 G} \frac{\Gamma}{\Delta} T_R \, \frac{U_{\rm max}}
	{\hbar} \, ,
\end{equation}
which is of the order of one recoil temperature per coherence time.
For our 100\,$\mu$K depth trap at $\Delta = 15\,$nm the rate is
4.4\,$\mu$K\,s$^{-1}$,
implying that storage and transport for many seconds is possible.
For simplicity, we have ignored the fact that there
may be distinct longitudinal and transverse temperatures which do not
equilibrate over the trapping timescales.

\subsection{Effect of a realistic substrate}
\label{sec:sub}

In this section we present calculations, performed using the method
of Section \ref{sec:num}, for a practical
substrate choice
of sodium fluoride (the lowest refractive index common mineral, with
$n_s = 1.32$ at a wavelength of 852\,nm), and investigate how this
changes the atom waveguide properties from those presented above.
We increased $n_g$ to 1.7 (dense flint glass, e.g. BaSF type)
in order to provide
sufficient index difference from the substrate.

Fixing the width at
$W = 1.00 \,\lambda$, we found that a height
$H = 0.34 \,\lambda$ gave the largest $\alpha_L$
of $0.20 \pm 0.01$, and allowed both modes to be
sufficiently far from cut-off (greater than half the power being carried inside
the guide for both modes).
The result is a goodness factor $G$ which is approximately half that of the
$n_s = 1$ case, with a
corresponding halving of the achievable product of depth and
coherence time according to (\ref{eq:expt}), and doubling of the
heating rate at a given $U_{\rm max}$ and $\Delta$ according to
(\ref{eq:tdot}).
The shorter decay lengths of 56\,nm and 68\,nm (compared to 93\,nm and 137\,nm
for $n_s = 1$) cause the typical trapping distance $y_0$ to be
reduced by a factor of roughly 1.8.

We found that in order to reproduce the depth of $100\,\mu$K
and $y_0 = 0.24\,\mu$m of the upper trap of Figure \ref{fig:pot}
(with $\Delta$ unchanged) we needed
$P_{\rm tot} = 22$\,mW, giving $\tau_{\rm coh} = 9\,$ms.
The large power increase over the 1\.mW required for $n_s
= 1$ is explained by the fact that this $y_0$ is now towards the upper
limit practically achievable rather than the lower.
(If $y_0$ is instead scaled in proportion to the new decay lengths,
the required increase in $P_{\rm tot}$ is only a factor 1.7).
In this example, we find the transverse oscillation frequencies have
increased to
$\omega_x / 2\pi = 81$\,kHz and
$\omega_y / 2\pi = 202$\,kHz, compared to the original
$\omega_x / 2\pi = 26$\,kHz and
$\omega_y / 2\pi = 109$\,kHz.
The increase in $\omega_y$ is explained entirely by the shorter decay lengths,
and the increase in $\omega_x$ (by a factor of over 3) is attributed to tighter
optical mode shapes.
It is clear that this latter effect outweighs the decrease in $\tau_{\rm coh}$,
implying that the inclusion of the substrate has
actually {\em increased} $Q$ by 50\%.

In summary, the
effects of including a realistic substrate limit the maximum trapping
distance $y_0$ that can be achieved (because of the reduction in decay lengths),
lower the goodness factor, increase the heating rate and the
required optical power, but also
increase the oscillation frequencies.
For our substrate choice, each of these changes was approximately
a factor of 2, and we believe that they do not alter the basic practicality
of implementing our proposed waveguide.

\section{Numerical solution of the light fields}
\label{sec:num}

The detailed electric field distribution is very important in calculating the
trapping potential above the waveguide. An approximation to the form
of the fields in the y direction is given
by the analytically-known solution for the slab waveguide,
but to get more accuracy and
knowledge of the full potential shape in the x-y plane,
we used a full-vector finite element
calculation.

The technique represents the electric and magnetic fields as
simple piecewise functions over many ``elements''
(regions) subdividing a slice through the guide and surrounding media
in the xy plane,
therefore by a finite number of degrees of freedom.
Each element has a dielectric constant associated with it, allowing arbitrary
stepwise refractive index distributions in the xy plane to be modelled.
Maxwell's equations for propagating solutions of the form
$\exp(ik_z z - i\omega t)$ are reduced to a generalized sparse
eigenvalue equation with $k_z^2$
as the eigenvalue and the bound mode field distributions as the eigenvectors
\cite{FEreview}.
Specifically, we used the technique of Fernandez and Lu
\cite{FL93},
with $H_x$
and $H_y$ as the field degrees of freedom, for simplicity
using first-order (bilinear) functions to represent these fields over a
non-uniform but separable rectangular grid of elements.
This required a generalization of the Fernandez and Lu implementation, and
careful consideration of their line-integral terms (which are non-standard
for a finite element formulation)
\cite{future}.
Of the many available finite element approaches to dielectric waveguide mode
solving, this frequency-domain method was chosen for its
absence of `spurious modes', its ability
to handle index step discontinuities, its small number of required
degrees of freedom and its matrix sparsity\cite{FLtest}.

Rather than emulating a radiative boundary condition
(a notoriously hard task usually requiring an iterative
procedure due to the $k_z$ dependence),
we enclosed the problem in a large, perfectly-conducting box
of sufficient
size
that the bound mode evanescent fields were negligible on its walls, making the
nature of the boundary condition irrelevant.
However,
the average level spacing of the unbound modes (the `continuum')
decreases with increasing box size,
and especially near cut-off this increases the number
of iterations required to solve the eigenvalue problem
to a given accuracy (we used the well-known ARPACK solver to find the 11 lowest
eigenmodes of the sparse matrix).
We found that a box size of $6\lambda$ to $7\lambda$ gave the best compromise
between accuracy and speed.
Our non-uniform elements allowed us to have a high element density across
the waveguide and in the trapping region, but
a low density over the much larger box area,
keeping the total number of degrees of freedom manageable.

\begin{figure}[t]
\centerline{\epsfig{figure=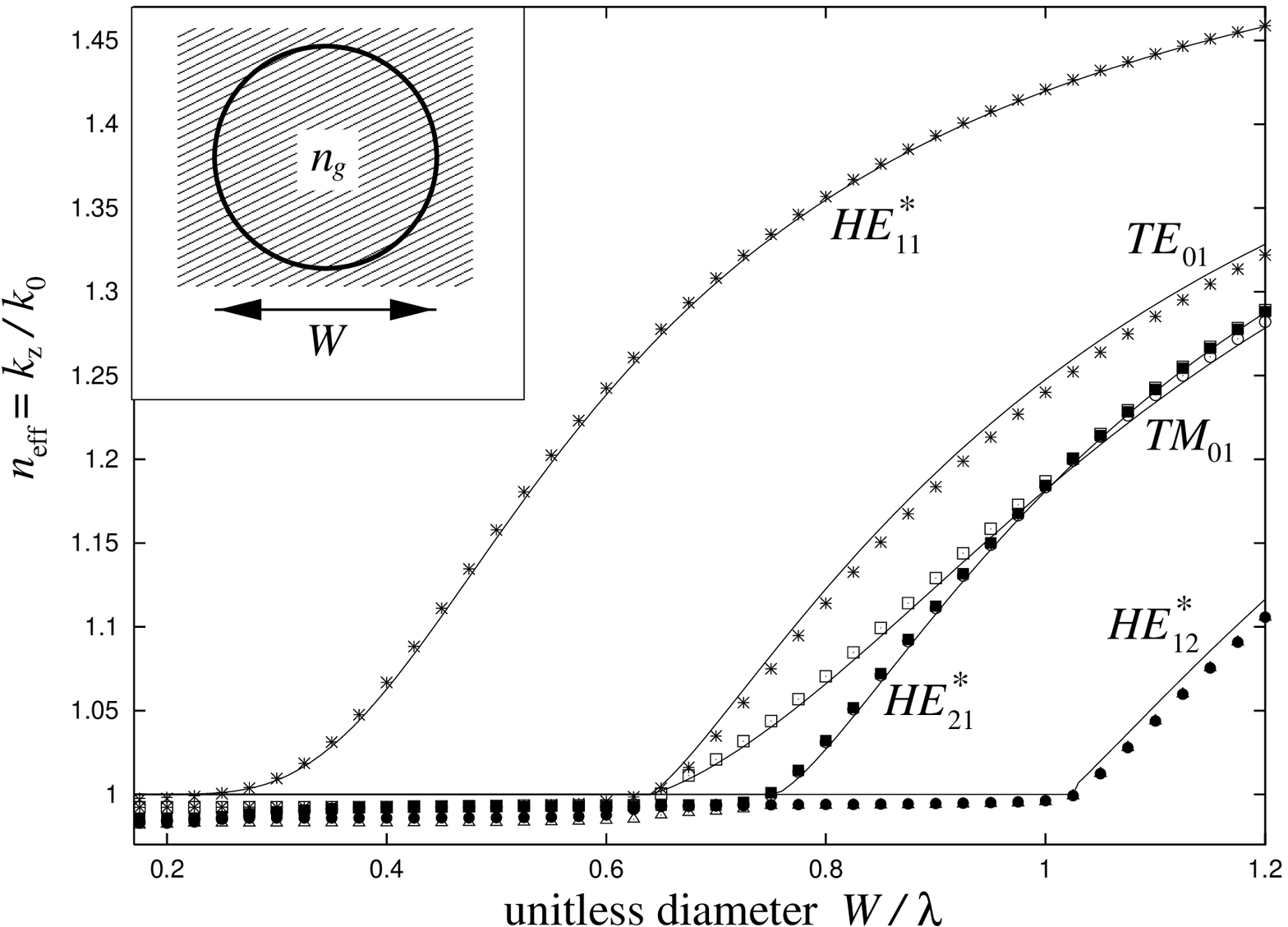,width=3.2in}}
\vspace{0.1in}
\caption{Comparison of our optical guide bound-mode numerical implementation
against known analytic solutions, in the case of a free-standing dielectric
cylinder
of $n_g = 1.56$.
We used discretization and box-size identical to the rectangular
guide case, and observe typical errors of $\pm 0.5\%$ in propagation constant
for the first two modes.
The mode naming convention and analytic calculation follow
Snitzer~\protect\cite{snitzer};
an asterisk indicates a doubly-degenerate mode.}
\label{fig:num}
\end{figure}

The fractional error $\epsilon$ in the propagation constant $k_z$ was
less than $1\%$, and the accuracy of the electric field strengths in the
trapping region $\approx 3\%$, when we used $N \sim 2000$ elements.
This was sufficiently accurate for the present work.
Finding the bound modes of each waveguide parameter choice
typically took between 3 and 20 minutes of computing time on a
Silicon Graphics R8000
processor, depending on how close to cut-off the guide was.
We tested the accuracy of the method by solving a cylindrical guide in an
identical fashion with the same $N$ and a very similar non-uniform grid, for
which there are
known field solutions \cite{snitzer}.
Figure \ref{fig:num} shows the propagation constant agrees with the analytics
to within 1\%, even
close to cut-off.
The convergence with $N$ was measured for the rectangular guide case, and
found to be
$\epsilon \sim N^{-\gamma}$ with $0.55 < \gamma < 0.7$.
This is less than
optimal for first-order elements (which have a maximum possible
convergence of $\gamma = 1$), and is believed to be
due to an inability of the bilinear functions to represent
physical in-plane $E$ and $H$ components at dielectric steps, or the weak
field singularities which can physically occur at any exterior dielectric
sharp edges (regardless of whether acute or obtuse)\cite{future}.

Future improvements to the method, which would increase the accuracy or the
convergence rate $\gamma$, include using
higher order elements (if done carefully, this could
correctly represent physical $E$ and $H$ components at dielectric steps), and
explicit modelling of the field singularities at
guide corners using specialized elements.

\section{Further decoherence and loss mechanisms}
\label{sec:further}

\subsection{Effects of surface interactions}
\label{sec:surf}

The EW trap has the benefit of creating high
field gradients near a surface, but along with this comes the disadvantage
that interactions with that surface that can alter the trap potential
and even cause heating
and loss of trapped atoms.

An atom's change in potential near a surface is
known as the van der Waals interaction ($l \ll \lambda$) or the Casimir
interaction ($l \gg \lambda$), depending on the distance $l$ from the surface
compared to 
$\lambda$, the dominant wavelength responsible for the
polarizability of the atom (in our case of Cs this is the D line resonance,
the same as our trapping resonance).
There is a smooth
cross-over from van der Waals ($U \sim l^{-3}$, which can be viewed
as the atom's electrostatic interaction with the 
image of its own fluctuating dipole) to Casimir ($U \sim l^{-4}$,
which can be viewed as a retarded van der Waals attraction
or equally well as an atomic level
shift due to a cavity QED effect) at
$l \approx \lambda/10$\cite{cavqed}.
In the case of a perfect mirror surface, the full form is known for any $l$,
but for a dielectric surface, the expression becomes much more
complicated to evaluate\cite{spruch}.

Since our trapping distances are larger than this cross-over point, we will use
the Casimir form, which is correct for asymptotically large $l$, and
is always an overestimate of the true potential\cite{cavqed}.
The dependence of the coefficient with dielectric constant is complicated
\cite{lifshitz,yan}, but
we will use the simpler approximate
form given by Spruch and Tikochinsky\cite{spruch}, to give
\begin{equation}
\label{eq:cas}
	U_{\rm Cas}(l) = -\frac{3}{8 \pi}
	\frac{\hbar c \,\alpha(0)}{4 \pi \epsilon_0 \, l^4} \,
	\frac{\epsilon - 1}{\epsilon+ (30/23)\epsilon^{1/2} + 7/23}
\end{equation}
(in the MKSA system).
This approximate form is known to be within 6\% of the
exact expression for any dielectric
constant $\epsilon$ \cite{yan}.
Substituting the recently calculated\cite{babb}
static polarizability of cesium, $\alpha(0) = 399.9$ a.u., gives a Casimir
interaction coefficient of 4.9\,nK\,$\mu$m$^4$ for $n_g = 1.56$.

Figure \ref{fig:geom}b shows the effect
of this potential on a
typical trap of depth 150\,$\mu$K and distance $y_0 = 270$\,nm.
It is clear that the change is negligible further than 100\,nm from the
surface,
and a WKB
tunneling calculation along this straight-line path (at $x=0$) shows that
even if all atoms that reach the surface stick, the loss rate from the
first few transverse modes is entirely negligible.
However, care should be taken with the multi-mode regime, or in the case of
high-$p$ traps, since the tunneling via the
corners of the ``bean'' shape may dominate for $p>0.4$ (Figure \ref{fig:pot}).

The issue of energy transfer to trapped atoms due to a finite (and possibly
room) temperature nearby surface is far less well
understood, and may be a problem
with many surface-based particle traps, as discussed by Henkel and Wilkens
\cite{henkel}.
However, since we are trapping neutral particles and the conductivity of our
surface is low, we expect a decoherence rate negligible compared
to that already present from spontaneous absorption and emission cycles.

\subsection{Bending the waveguide}
\label{sec:bend}

It would be very useful to be able to carry atom beams
along curved paths, by bending our
atom waveguide in the plane of the substrate, without significant atom
loss.
Here we briefly estimate three limitations on the waveguide minimum bending
radius (in decreasing order of leniency):
1) the limit imposed by optical radiation leakage,
2) the limit needed for incoherent atom transport,
and 3) the limit needed
for coherent atom transport in the transverse groundstate.
This will give us an idea of the practicality of curved atomic guides.

Firstly, whenever a dielectric optical guide has curvature, there is a loss
rate (exponentially small in the curvature radius $R$), which can
be viewed as tunneling out of the guide's `potential well' induced by
the addition of an effective centrifugal potential.
In the limit $R \gg W$, the effective potential is linear with x (the radial
coordinate), and the fractional loss per radian of curvature can be
estimated\cite{bellsystems},
for instance using the one-dimensional WKB formula, to be
\begin{equation}
\label{eq:wkb}
	\alpha = C \, \frac{R}{\lambda} \exp \left( -\frac{1}{6 \pi^2}
	\frac{\lambda^2 R}{L_x^3}
	\right) \, ,
\end{equation}
where $L_x$ is the evanescent decay length in the radial direction, and $C$ is
a constant of order unity.
Therefore for negligible light loss at a $\pi/2$ bend we need
$R > 60\pi^2 L_x^3 / \lambda^2$, typically a couple of tens of microns.
This is so small chiefly because we are using an optical guide with
a large refractive index
step\cite{tamir}.

Secondly, we consider atom
loss from an incoherent beam with a
transverse temperature $k_B T_\bot$ (assumed small compared to the
trap depth magnitude $U_{\rm max}$),
and a longitudinal
kinetic energy $E_\|$.
We call the approximate spatial
extent of the trap potential in the x direction $2 \xi$,
and restrict ourselves to one-dimensional classical
motion in this direction.
When in a region of radius of curvature $R \gg \xi$,
an effective centrifugal term adds to the
trapping potential giving
$U(x) = U_{\rm dip}(x) - 2E_\| (x/R)$.
This causes the atoms to `slosh' towards positive x, only ever returning
if there
exists a point where $U(x) > -U_{\rm max}$ for $x > 0$.
We can estimate that this will happen if $U_{\rm max} > 2E_\| (\xi/R)$,
giving our lower
limit on $R$ as $2\xi E_\| / U_{\rm max}$.
In our design $\xi \approx 0.5\,\mu$m, so if we choose $R = 1$\,mm we can
expect loss-free transport of a beam at a longitudinal kinetic
energy up to $10^3$ times the trap depth.

Thirdly, to model coherent matter-wave propagation along a curved guide,
we consider the amplitude for remaining in the transverse ground-state, having
passed into a curved section and back into a straight section.
If again we assume one-dimensional x motion, and assume a harmonic potential
$U(x) = \smallfrac{1}{2} M \omega_x^2 (x-x_0)^2$ around the trap minimum, then
the effect of curvature is to shift the minimum position from $x_0 = 0$
to $x_0 = 2 E_\| / M \omega_x^2 R$.
If this shift is much less than the characteristic ground-state size
$(\hbar/M \omega_x)^{1/2}$ then the projection at each transition will be high,
resulting in high flux transmission coefficient.
This gives $R \gg 2 E_\| / (\hbar M \omega_x^3)^{1/2}$ as our condition,
which for $E_\| / k_B = 10\,$mK (that is, $v_\| = 1.1$\,ms$^{-1}$)
and $\omega_x/2\pi = 40$\,kHz
corresponds to $R \gg 0.45$\,mm.
This limit is very conservative since we have not yet made
use of the adiabatic condition
$\Omega \ll \omega_x$ (where $\Omega \equiv v_\| /R$ is the rate of
change of direction of the guided atom),
to design a waveguide path without discontinuities in
the curvature.

In conclusion, we have shown that it is possible to bend atoms
both incoherently and
coherently through large angles on a compact substrate structure of a few
millimeters in size.

\section{Conclusion}
\label{sec:conc}

We have presented a novel substrate-based neutral atom 
waveguide, based on the optical dipole force,
which combines the features of a planar far-detuned two-color evanescent
trap\cite{ov91} with the ability
to confine strongly along two axes.
We utilized
the differing vertical decay lengths
of the two
bound-mode {\em polarizations}
of a submicron-sized optical waveguide near cut-off.
We have shown that only a few milliwatts of guided laser power
can create atomic potential depths $\sim 100\,\mu$K with
transverse oscillation frequencies $\sim 100\,$kHz,
a coherence time
$\sim 10\,$ms and a trap minimum 200-400\,nm above the optical guide surface,
for Cs atoms in the $m_F = 0$ state.
Laser powers greater than ten milliwatts
can give a transverse mode spacing greater than
the temperature of a Cs MOT, opening up the single-mode waveguide regime.
The advantages of guiding optical trapping fields on a substrate include 
mechanical stability and reliability, mass production and the potential for
transport along complicated paths.

We have given
some design criteria for guided-lightwave two-color
atom waveguides (chiefly the maximization of the evanescent
decay lengths, and of their normalized difference $\alpha_L$),
and shown that a substrate of low refractive
index can be very beneficial.
We modelled in detail the trapping potentials
for a general rectangular guide of index 1.56
above a unity-index substrate, and
have shown that a realistic substrate choice of index
1.32 poses few problems to the viability of the device.
We predict that the
effect of the surface interaction is generally small, and that
coherent guiding is possible around corners of radii $\sim 1\,$mm for a
longitudinal velocity $\sim 1\,$ms$^{-1}$.
We also believe that the magnetic part of the potential
felt by nonzero $m_F$ atoms could be used to increase the depth and oscillation
frequencies further.

This preliminary work (specifically equation (\ref{eq:tradeoff})) indicates
that utilizing detunings much larger than the 15\,nm we
limit ouselves to
here will be very advantageous for coherent guiding.
We have only scratched the surface of the design variations
possible; for instance, equalizing the horizontal
and vertical oscillation frequencies is yet to be attempted.
The use of two polarizations is our solution to the problem
of maximizing $\alpha_L$ when the detuning is very small compared to the
wavelength, but we suspect that there will exist other fruitful schemes
where these are comparable ($\Delta \sim \omega_0$, very far-detuning)
and where a large $\alpha_L$
is caused simply by the different optical
cut-off conditions at the two wavelengths.
We have reserved investigation of cooling schemes for future work (although
this has already been demonstrated in an EW mirror\cite{desbiolles}
and proposed
in EW traps\cite{loading}).
We believe that the
potential shapes capable of being produced by guided waves on a substrate
also include the possibility of funnel-type loading schemes and
coherent atom couplers, allowing for a complete ``integrated''
atom-optical experiment on a substrate.

\acknowledgements
It our pleasure to thank J.~Thywissen, J.~Babb, N.~Dekker, V.~Lorent,
and the members of the
Heller Group for fruitful discussions.
AHB would also like to thank Prof E.~J.~Heller for computing resources.
This work was supported by the National Science Foundation, on
grant numbers CHE-9610501 and PHY-9732449.



\begin{figure}[t]
\centerline{\epsfig{figure=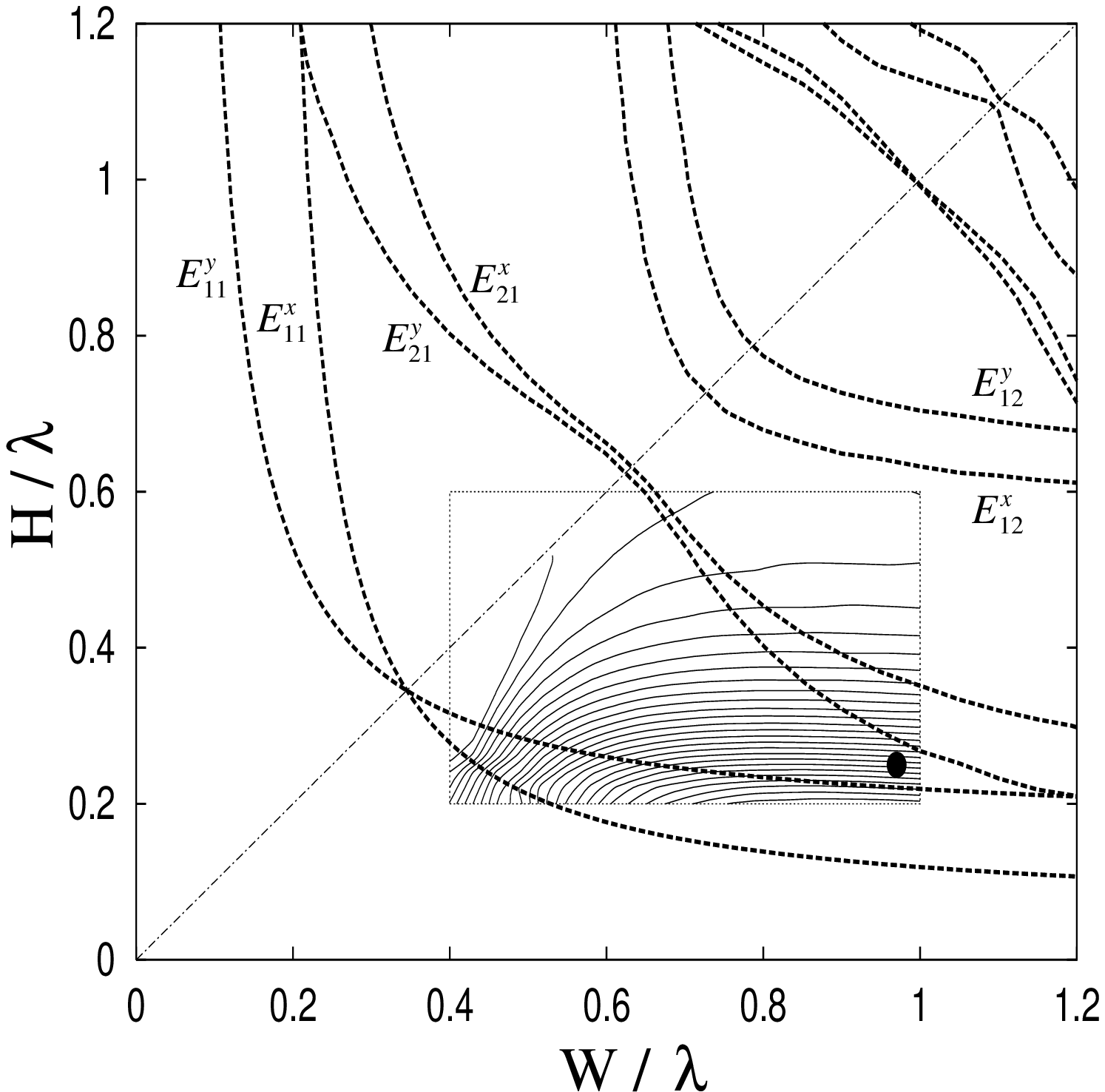,width=3.2in}}
\vspace{0.1in}
\caption{Numerically solved cut-off curves for a dielectric waveguide
of $n_g = 1.56$ (with $n_s = 1$)
as a function of its
width and height (dashed curves)
and, on the same axes, contours of the maximum trapping potential depth
achievable at fixed total laser power
(thin solid curves in rectangular overlayed box region).
Also shown is the symmetry line $W=H$
(thin dash-dotted line).
Note that the contours show depth {\em increasing} as $H$ decreases,
almost independent of
$W$, near the suggested
operating dimensions (shown as a solid ellipse).
Cut-off is defined as reaching
an effective refractive index $n_{\rm eff} \equiv
k_z/k_o = 1.05$, except for $E^x_{21}$ and $E^y_{21}$ (thick dashed lines)
which we show cut-off at
$n_{\rm eff} = 1.02$.}
\label{fig:cutoff}
\end{figure}

\begin{references}
\bibitem{FLnote}
Following the notation of Fernandez \&
Lu\cite{FL93} for rectangular guides,
mode $E_{nm}^k$ has $n$($m$) antinodes in the $x$($y$) direction,
and overall polarization axis $k=x,y$.
However as the symmetry line $W=H$ is approached
(see Figure \ref{fig:cutoff}),
mixing of polarizations occurs for the case $n \ne m$
and the notation becomes more arbitrary.
\bibitem{freqnote}
Even though the frequencies will actually
differ
by a couple of percent because of
their detunings, and cause a difference in decay lengths of the correct sign,
this effect alone is not large enough to create large
$\alpha_L$ values. Future designs with larger detunings, possibly much larger
than the fine splitting,
may be able to
use this effect alone.
\bibitem{detuningnote}
We made the assumption that the positive and negative
detunings caused negligible shifts in the wavelength used for
the bound mode calculation. For the
largest detunings used in this paper the fractional
shift is 1.8\%.
\end{references}
\end{document}